\documentclass[aps,prl,twocolumn,showkey,square,numbers,amsthm,amssymb,amsmath,nobibnotes,footinbib]{revtex4}  
\usepackage{esint}
\usepackage{bm,bbm}  
\usepackage{eucal}
\usepackage{verbatim}
\usepackage{booktabs,array}
\usepackage{times}
\usepackage{graphicx,color}
\usepackage{tabularx}
\usepackage{dsfont}
\usepackage{titletoc} 
\usepackage{microtype}
\usepackage[colorlinks=true,linkcolor=black,citecolor=black,urlcolor=blue]{hyperref}

\newcommand{\barr}{\begin{eqnarray}}
\newcommand{\earr}{\end{eqnarray}}
\newcommand{\be}{\begin{equation}}
\newcommand{\ee}{\end{equation}}

\def\Ord{\mathcal{O}}
\newcommand{\de}{\mathrm{d}}

\newcommand{\numberset}{\mathbb}

\newcommand{\R}{\numberset{R}}
\newcommand{\C}{\numberset{C}}

\newcommand{\avg}[1]{\left< #1 \right>} 

\def\im{{\rm i}}

\newcommand{\ZZ}{\mathcal{Z}}

\begin{document}

\title{Fluctuations in the two-dimensional one-component plasma and \\ associated fourth-order phase transition}

\author{Fabio Deelan Cunden$^{1,2}$, Anna Maltsev$^{1}$, Francesco Mezzadri$^{1}$}
\address{\mbox{$1.$ School of Mathematics, University of Bristol, University Walk, Bristol BS8 1TW, United Kingdom}\\
\mbox{$2.$ Istituto Nazionale di Fisica Nucleare (INFN), Sezione di Bari, I-70126 Bari, Italy}
}

\begin{abstract}
We study the distribution of the mean radial displacement of charges of a 2D one-component plasma in the thermodynamic limit $N\to\infty$ at finite temperature $\beta>0$. We compute explicitly the large deviation functions showing the emergence of a fourth-order phase transition as a consequence of a change of topology in the plasma distribution. This weak phase transition occurs exactly at the ground state of the plasma. These results have been compared with the integrable case (finite $N$) of plasma parameter $\beta q^2=2$. In this case the problem can be mapped to the stationary properties of 2D Dyson Brownian particles and to a non-Hermitian matrix model.
\end{abstract}

\maketitle

\textit{Introduction.} The two-dimensional one-component plasma (2D-OCP) is one of the most basic statistical mechanics fluid (i.e. non lattice) models in dimension greater than one. Very recently, Sandier and Serfaty~\cite{Sandier} have rigorously established a connection between the Ginzburg-Landau model~\cite{GL} for superconductors in the critical regime with vortices and the 2D-OCP. In this regime the interaction between vortices is well-described by a one-component plasma in a quadratic confining potential. Heuristic arguments suggest that the very same approximation could be valid to describe more general vortex systems, most notably in superfluids or Bose-Einstein
condensates~\cite{Correggi08,Correggi12}.

The 2D-OCP (also known as \emph{log-gas}) is a system of classical pointlike particles
of same charge $q$ (one species of particles) immersed in a two-dimensional domain with a neutralising background of opposite charge. The interaction between the charges is $(1/2)\sum_{i\neq j}q^2v(|\vec{r}_i-\vec{r_j}|)$, where the Coulomb potential is solution of the 2D Poisson equation, $v(\vec{r})=-\log(|\vec{r}|/L)$. The Boltzmann-Gibbs canonical measure of a 2D-OCP of $N$ particles at inverse temperature $\beta$ in a quadratic potential is
\begin{align}
\mathbb{P}_{\beta}\left(\left\{\vec{r}_k\right\}\right)&=\frac{1}{\mathcal{Z}_{N,\beta}}e^{-\beta H(\vec{r}_1,\dots,\vec{r}_N)} \label{eq:jpdf}\\
H\left(\left\{\vec{r}_k\right\}\right)&=-\frac{q^2}{2}\sum_{i\neq j}\log\left(\frac{r_{ij}}{L}\right)+\frac{q^2N}{2}\sum_k\left(\frac{r_k}{L}\right)^2, \label{eq:hamiltonian}
\end{align}
where $\vec{r}_k=(x_k,y_k)\in\R^2$ is the position of the $k$-th particle ($k=1,...,N$), $r_k=|\vec{r_k}|$ and $r_{ij}=|\vec{r}_i-\vec{r}_j|$.  The  thermodynamic
state of the system is characterised essentially by the two-dimensional plasma parameter
$\gamma =\beta q^2$. Hereafter the length $L$ that fixes the zero of the potential and the elementary charge $q$ will be scaled to unity. (The plasma parameter $\gamma$ is therefore identified with the inverse temperature $\beta$.) 

Besides being an approximation of the Ginzburg-Landau model in the vortices phase, these log-gases are interesting \emph{per se}, being ubiquitous in many fields of physics and mathematics. At the special value $\beta=2$ of the plasma parameter, the model is exactly integrable and deserves particular attention. In this case, \eqref{eq:jpdf} is the stationary distribution of a 2D Dyson Brownian motion~\cite{Osada13,Dyson62}, namely a Brownian motion of $N$ particles with logarithmic repulsion in the plane confined in a harmonic potential. Again at $\beta=2$, with the identification $\R^2\simeq\C$,  \eqref{eq:jpdf} is the joint distribution of the eigenvalues $\lambda_k=x_k+\im y_k$ of non-Hermitian random matrices with iid complex Gaussian entries, the Ginibre ensemble~\cite{Ginibre}.  There has been extensive work on 2D-OCP systems in the mean-field theory; numerical studies are also available, while exact results can be derived in the integrable case $\beta=2$~\cite{Deutsch79,Jancovici81,Alastuey81,Johannesen83,Forrester85,Cornu88,Forrester98,Samaj03,Fischmann11,Choquard83}. 

In this paper we present a study on the large deviations (rare events) of 2D-OCP systems. We will consider a rather simple observable, namely the mean radial displacement $\Delta_N=N^{-1}\sum_{k}r_k$ of the charges of the plasma (the vortices of the Ginzburg-Landau model in the critical phase). Here we compute explicitly the large $N$ limit probability law $\mathcal{P}_{N,\beta}(x)=\avg{\delta(x-\Delta_N)}$ of $\Delta_N$ and its rescaled Laplace transform  $\widehat{\mathcal{P}}_{N,\beta}(s)=\langle e^{-\beta sN^2\Delta_N}\rangle$ (see equations~\eqref{eq:J}-\eqref{eq:result_J} and~\eqref{eq:rate_funct_def}-\eqref{eq:rate_func_asym} below) for all $\beta>0$ (the angle brackets denote the canonical average). As a byproduct, at $\beta=2$ these results provide large deviations functions for the Ginibre ensemble and 2D Dyson Brownian particles. 

Results on the probability of rare events are popular in the one-dimensional case, in particular in the literature on Hermitian random matrices~\cite{Dean06,Facchi08,Vivo08,DePasquale10,Majumdar11,Borot12,Facchi13,Texier13}. In most of the models analysed the 2D-OCP is confined on the line and undergoes a phase transition, usually of third-order if the plasma is constrained by a hard wall (for an interesting review see~\cite{Majumdar14}). Large deviations have been much less explored in two dimensions and few results are available \cite{Jancovici93,Allez14,Arous98,Petz98}. Here we show explicitly that the statistical distribution of the mean displacement $\Delta_N$ in the plane unveils a \emph{fourth-order phase transition} of the 2D-OCP related to a change of topology in the equilibrium distribution of the plasma. This phenomenon is absent in one dimension. We point out that the extraordinary weak singularity of $\mathcal{P}_{N,\beta}(x)$ (or $\widehat{\mathcal{P}}_{N,\beta}(s)$) occurs at the average value $x=\avg{\Delta_N}$ (at $s=0$ in the Laplace domain), i.e. at the ground state of the 2D plasma. Therefore this phase transition may be detected by the anomalous behaviour of the high order cumulants of $\Delta_N$.

\textit{Results.} The general theory of large deviations of log-gases~\cite{Arous98,Petz98} ensures that the Laplace transform $\widehat{\mathcal{P}}_{N,\beta}(s)$ of $\Delta_N$ has a thermodynamic limit
\be
J(s)=-\lim_{N\to\infty}\frac{1}{\beta N^2}\log\widehat{\mathcal{P}}_{N,\beta}(s) \label{eq:J}
\ee
for all $s\in\R$. Such a limit is the scaled \emph{cumulant generating function} (cumulant GF) of $\Delta_N$. Then, by the G\"{a}rtner-Ellis theorem~\cite{Gartner,Ellisthm}, the probability law of $\Delta_N$ satisfies a large deviation principle~\footnote{For more references on the terminology see~\cite{Ellis99,Touchette09}.} with \emph{speed} $\beta N^2$, i.e.  \mbox{$\mathcal{P}_{N,\beta}(x)\approx e^{-\beta N^2\Psi(x)}$}, where the \emph{rate function} $\Psi(x)$ is the Legendre-Fenchel transform of the cumulant GF~\eqref{eq:J}:
\be
\Psi(x)=-\lim_{N\to\infty}\frac{1}{\beta N^2}\log\mathcal{P}_{N,\beta}(x)=\inf_s\left[J(s)-sx\right]. \label{eq:rate_funct_def}
\ee
\begin{figure}[t]
\centering
\includegraphics[width=1\columnwidth]{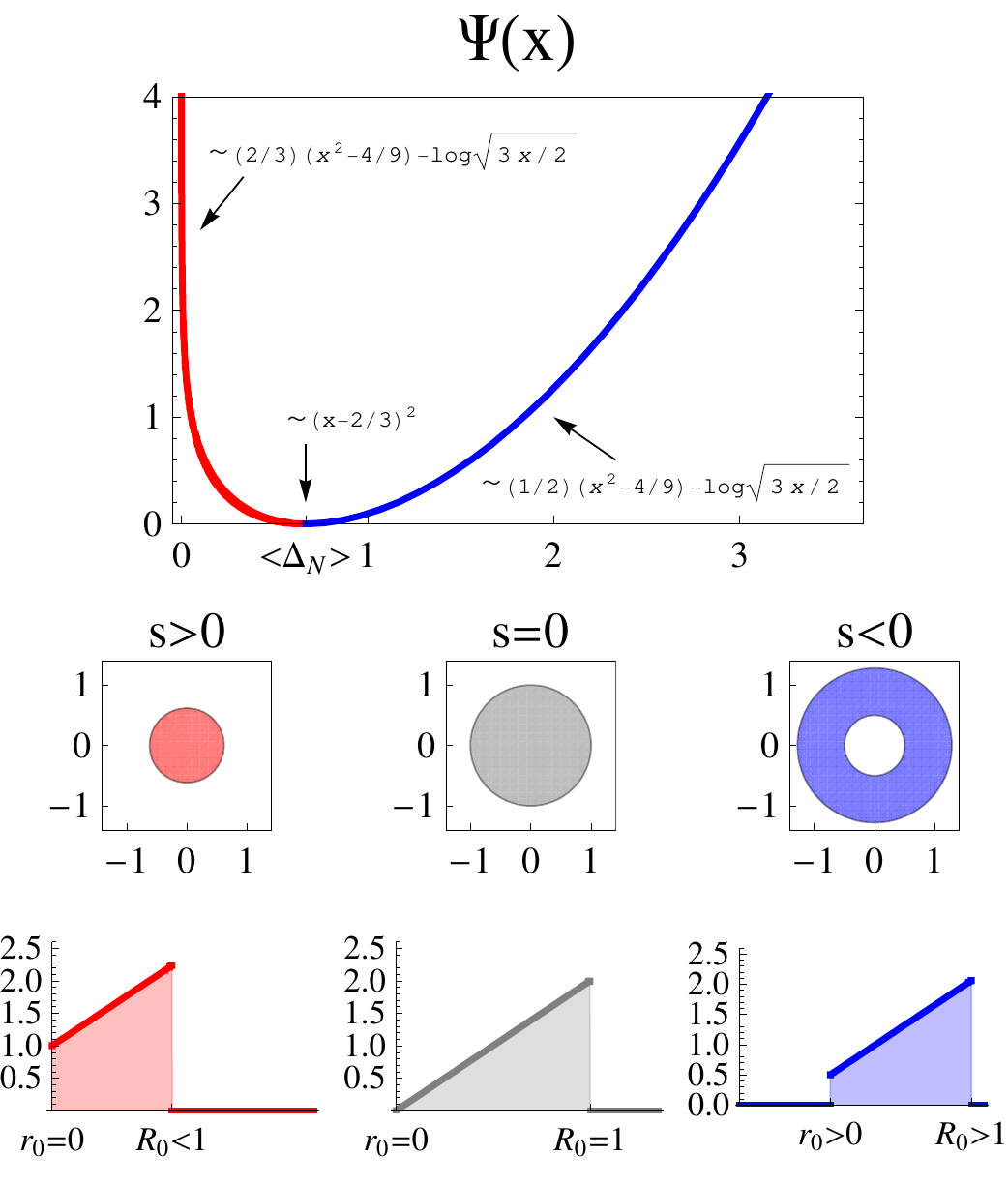}
\caption{(color online) Top: Rate function $\Psi(x)$ from~\eqref{eq:result_Psi} and its leading terms for $x\to0$, $x\simeq2/3$ and $x\to\infty$ (see~\eqref{eq:rate_func_asym}). At $x=\avg{\Delta_N}\equiv2/3$ the function is not analytic. Central: Support $\Sigma(s)$ of the equilibrium distribution $\mu^{\star}_s$ of the plasma for different values of $s$. From left to right $s=1,0,-0.5$. Bottom:  The corresponding radial distributions $\int_0^{2\pi}\de\mu^{\star}_s(|\vec{r}|,\varphi)$.}
\label{fig:rateF}
\end{figure}

The challenge here is to compute explicitly the large deviation functions $J(s)$ and $\Psi(x)$. We will show that the cumulant GF has the following expression independent of $\beta$:
\be
J(s)=\frac{1}{2}\mathrm{arsinh}\left(\frac{s}{2}\right)-\frac{s^2}{4}+\frac{s}{48}\left[\left(s^2+10\right)\sqrt{s^2+4}-|s|^3\right].
\label{eq:result_J}
\ee
This is the central result of the paper. Before turning to the derivation, we make few observations and we discuss the consequences of~\eqref{eq:result_J}. The cumulant GF is strictly concave and satisfies the normalization condition $J(0)=0$. Notice that the large deviation function is not analytic. The \emph{fourth} derivative of $J(s)$ is discontinuous at $s=0$ with a finite jump $J^{(4)}(0^-)-J^{(4)}(0^+)=1$; this is a consequence of a \emph{fourth-order phase transition} in the equilibrium distribution of the 2D-OCP. Phase transitions of 2D-OCP constrained on a line, i.e. of eigenvalues of Hermitian random matrices, are well understood. In two dimensions, however, they are more gentle than for Hermitian matrices, where the singularities are of third order. This behaviour corresponds to the physical intuition that the logarithmic repulsion has stronger effects in 1D than in 2D gases. Here the singularity emerges in the thermodynamic limit as a consequence of a change in topology (disk-annulus) of the support of the equilibrium distribution of the charges. This mechanism is absent in one dimension. We note that the phase transition would not occur in the absence the logarithmic interaction, since without it \eqref{eq:hamiltonian} is the Hamiltonian of a system of free particles, which cannot undergo a phase transition --- this case is discussed in more detail in the last section. As a final remark, we stress that the phase transition occurs \emph{at} the ground state ($s=0$) and hence is not related to the atypical fluctuations (this is another departure from the one-dimensional scenario for continuous statistics). 

At leading order in $N$, the first three cumulants of $\Delta_N$ can be obtained by differentiation of $J(s)$ at $s=0$
\be
\kappa_1(\Delta_N\!)=\frac{2}{3};\, \kappa_2(\Delta_N\!)=\frac{1}{2\beta N^2};\, \kappa_3(\Delta_N\!)=\frac{1}{2\beta^2N^4}.
\label{eq:cumulants}
\ee 
At $\beta=2$ the first two cumulants (average and variance) agree with asymptotic formulas for the Ginibre matrix ensemble~\cite{Forrester99}. The nonzero third cumulant measures the deviation from the Gaussian behaviour at leading order in $N$. Higher cumulants cannot be obtained by differentiation of $J(s)$ (we will elaborate more on this point later).
Since $J(s)$ is everywhere differentiable, we can apply the G\"{a}rtner-Ellis theorem and recover the rate function $\Psi(x)$ using~\eqref{eq:rate_funct_def}. The limiting behaviour of $\mathcal{P}_{N,\beta}(x)\approx e^{-\beta N^2\Psi(x)}$ may be summarized as follows
\begin{align}
\mathcal{P}_{N,\beta}(x)\approx
\begin{cases}
\left(\frac{2x}{3}\right)^{\beta N^2/2}e^{-(2/3)\beta N^2\left(x^2-4/9\right)}&\text{for $x\ll1$}\vspace{2mm}\\
\exp{\left[-\beta N^2\left(x-\frac{2}{3}\right)^2\right]}&\text{for $x\simeq\frac{2}{3}$}\vspace{2mm}\\
\left(\frac{2x}{3}\right)^{\beta N^2/2}e^{-(1/2)\beta N^2\left(x^2-4/9\right)}&\text{for $x\gg1$}.
\end{cases}\label{eq:rate_func_asym}
\end{align}
The rate function $\Psi(x)$ is plotted in the top panel of Fig.\ref{fig:rateF}. From the Gaussian form around $x\simeq2/3$, one easily reads off the values for mean and
variance: $\avg{\Delta_N}=2/3$ and $\mathrm{var}(\Delta_N\!)=1/(2\beta N^2)$, according to~\eqref{eq:cumulants}. On the other hand, the rate function is not quadratic and this implies the polynomial and sub-Gaussian far tails of $\mathcal{P}_{N,\beta}(x)$  for $x\to0$ and $x\to\infty$, respectively, in~\eqref{eq:rate_func_asym}. 

We stress that the exact computation of $J(s)$ is essential to unveil the phase transition. Standard computations usually concern with the quadratic approximation about the ground state of the 2D-OCP (Gaussian approximation of $J(s)$ about $s=0$). However the fourth-order singularity at $s=0$ does not emerge at the level of quadratic (i.e. $O(s^2)$) effects.

\textit{Derivation.} In order to compute the cumulant GF~\eqref{eq:J}, we cast the Laplace transform of $\Delta_N$ as a ratio of two partition functions $\widehat{\mathcal{P}}_{N,\beta}(s)=\frac{\ZZ_{N,\beta}(s)}{\ZZ_{N,\beta}(0)}$, where $\ZZ_{N,\beta}(s)=\int e^{- \beta H(\vec{r}_1,\dots,\vec{r}_N;s)}$ is the partition function of the 2D-OCP with modified energy
\begin{align}
H(\vec{r}_1,\dots,\vec{r}_N;s)&=-\frac{1}{2}\sum_{i\neq j}\log r_{ij}+N\sum_kV_s(r_k) \label{eq:hamiltonian_s}\\
V_s(r)&=\frac{1}{2}r^2+sr\ . \label{eq:potential_s}
\end{align}
Clearly, at zero temperature $\beta\to\infty$ the only contribution to $\ZZ_{N,\beta}(s)$ comes the configurations of the system that minimize the energy $H(\vec{r}_1,\dots,\vec{r}_N;s)$. At finite temperature $\beta<\infty$ in the large $N$ limit, the picture is not very different in the sense that the` typical' configurations of the system are still the minimizers of the energy and one can characterize the  probability of `atypical' configurations. For large $N$, we have  $H(\vec{r}_1,\dots,\vec{r}_N;s)=N^{2}(H_s[\mu_N]+o(1))$ where $\mu_N=N^{-1}\sum_i\delta_{\vec{r}_i}$ is the normalized distribution of the charges  and $H_s[\cdot]$ is the mean-field energy functional:
\be
H_s[\mu]=-\frac{1}{2}\iint\limits_{\vec{r} \neq \vec{r}'}\de\mu(\vec{r})\de\mu(\vec{r'})\log|\vec{r}-\vec{r'}|+\int\de\mu(\vec{r})V_s(|\vec{r}|). \label{eq:mean-field}
\ee
These considerations justify a saddle-point evaluation of the large deviation function $J(s)$ as excess free energy
\begin{align}
J(s)
&=-\lim_{N\to\infty}\frac{1}{\beta N^2}\log\left[\frac{\int e^{-\beta H(\vec{r}_1,\dots,\vec{r}_N;s)}}{\int e^{-\beta H(\vec{r}_1,\dots,\vec{r}_N;0)}}\right]\nonumber\\
&= H_s[\mu^{\star}_s]- H_0[\mu^{\star}_0],\label{eq:en_diff}
\end{align}
where $\mu^{\star}_s$ (resp. $\mu^{\star}_0$) is the minimizer of the energy functional $H_s[\cdot]$ (resp. $H_0[\cdot]$): 
\be
H_s[\mu^{\star}_s]=\min\left\{H_s[\mu]\colon\,\, \mu\geq0 ,\,\int_{\R^2}\de\mu(\vec{r})=1\right\}. \label{eq:min}
\ee
Hence, the problem of computing $J(s)$ reduces to find the equilibrium configuration $\mu^{\star}_s$ of the charges of plasma in the large $N$ limit.
Find the configuration of the 2D-OCP that minimizes a mean-field energy functional like~\eqref{eq:mean-field} is one of the central objectives in \emph{potential theory}~\cite{Saff}. 
Fortunately, minimization problems such as~\eqref{eq:min} can be solved in closed form with a judicious use of classical electrostatics. We first observe that the problem is radially symmetric and the radial potential $V_s$ is convex. Therefore, the equilibrium measure inherits the radial symmetry of the energy $\de\mu^{\star}_s(\vec{r})=(2\pi)^{-1}\de\mu^{\star}_s(r)\de\varphi$ and is supported in general on the annulus 
\be
\Sigma(s)=\left\{\vec{r}\in\R^2\colon\, r_0(s)\leq r \leq R_0(s)\right\},
\ee 
where we have switched to polar coordinates $\vec{r}=(r,\varphi)$. At equilibrium, the Gauss law on a generic domain  of the plane holds. For a centered disk of radius $r$, the Gauss law reads
\be
V_s'(r)r=2\pi\int_{0}^{r}\de\mu^{\star}_s(r')\,r'. \label{eq:Gauss_law}
\ee
Using the positivity and the normalization of $\mu^{\star}_s$, it is easy to find the conditions $V_s'(r_0)=0$ and $V_s'(R_0)R_0=1$ on the inner and outer radii of the support $\Sigma(s)$.
From~\eqref{eq:Gauss_law}, the equilibrium distribution of the charge has the explicit form
\begin{align}
\de\mu^{\star}_s(\vec{r})&=\left(\frac{2|\vec{r}|+s}{2\pi|\vec{r}|}\right)\,1_{\vec{r}\,\in\Sigma(s)}\,\de \vec{r},\label{eq:mus}\\
r_0(s)=\max&\left\{0,-s\right\},\quad R_0(s)=\frac{1}{2}\left[\sqrt{s^2+4}-s\right]. \label{eq:r0R0}
\end{align}
We see that the optimal distribution of the plasma experiences a change of topology driven by $s$: $\mu^{\star}_s$ is supported on a disk for $s\geq0$ and on an annulus for $s<0$ (see Fig.~\ref{fig:rateF}). The fourth-order singularity of $J(s)$ at $s=0$ witnesses this sudden disk-to-annulus change of topology. In particular, for $s=0$ we get the uniform measure on the unit disk: $\de\mu^{\star}_0(\vec{r})=\frac{1}{\pi}\,1_{|\vec{r}|\leq1}\de \vec{r}$. This corresponds to the \emph{circular law}~\cite{Ginibre,Girko84,Bai97,Bordenave12}, one of the most celebrated results on the empirical density of the eigenvalues of non-Hermitian random matrices. For a generic value $s\in\R$, the distribution $\mu^{\star}_s$ is the typical configuration of the 2D-OCP with a fixed value of $\Delta_N$ given by
\begin{align}
x(s)=\int\,\de\mu^{\star}_s(\vec{r}) |\vec{r}|
&=\frac{\left(s^2+4\right)^{3/2}-6s-|s|^3}{12}. \label{eq:x(s)}
\end{align}
The energy of the equilibrium configuration at $s=0$ (i.e. at the ground state) is $H_0[\mu^{\star}_0]=3/8$, a value known as $H$-stability bound of the 2D-OCP~\cite{Sari76,Sari76a}. For a generic value of $s$, the minimal energy attained at $\mu^{\star}_s$ is 
\be
H_s[\mu^{\star}_s]=\frac{1}{2}\left[\int\de\mu^{\star}_s(\vec{r})V_s(|\vec{r}|)+V_s (R_0)-\log R_0\right]. \label{eq:optimal_en}
\ee
Evaluating~\eqref{eq:optimal_en}, we get the cumulant GF $J(s)$ in~\eqref{eq:result_J} as excess energy~\eqref{eq:en_diff}.

Since $J(s)$ is strictly concave and differentiable, using the properties of the Legendre-Fenchel transform, we can write the thermodynamical relation $J(s)-\Psi(x)=sx$, where the conjugate variables $x$ and $s$ are related by  
\be
x(s)=J'(s)\quad\text{and}\quad s(x)=-\Psi'(x), \label{eq:thermo}
\ee
with $x(s)$ given in~\eqref{eq:x(s)}. These relations, supplemented with the conditions $J(0)=\Psi(x(0))=0$,
provide the following simple expression for the rate function $\Psi(x)$:
\be
\Psi(x)=-\int_{\frac{2}{3}}^{x} s(x')\,\de x', \label{eq:result_Psi}
\ee
where $s(x)$ is the (real non increasing) inverse of~\eqref{eq:x(s)} and we have used $x(0)=2/3$. One may verify that the cumulant GF~\eqref{eq:result_J} can be computed also from~\eqref{eq:thermo}. The asymptotic analysis of $\Psi(x)$ in~\eqref{eq:result_Psi} provides the limiting behaviour summarized in~\eqref{eq:rate_func_asym}.
\begin{figure}[t]
\centering
\includegraphics[width=1\columnwidth]{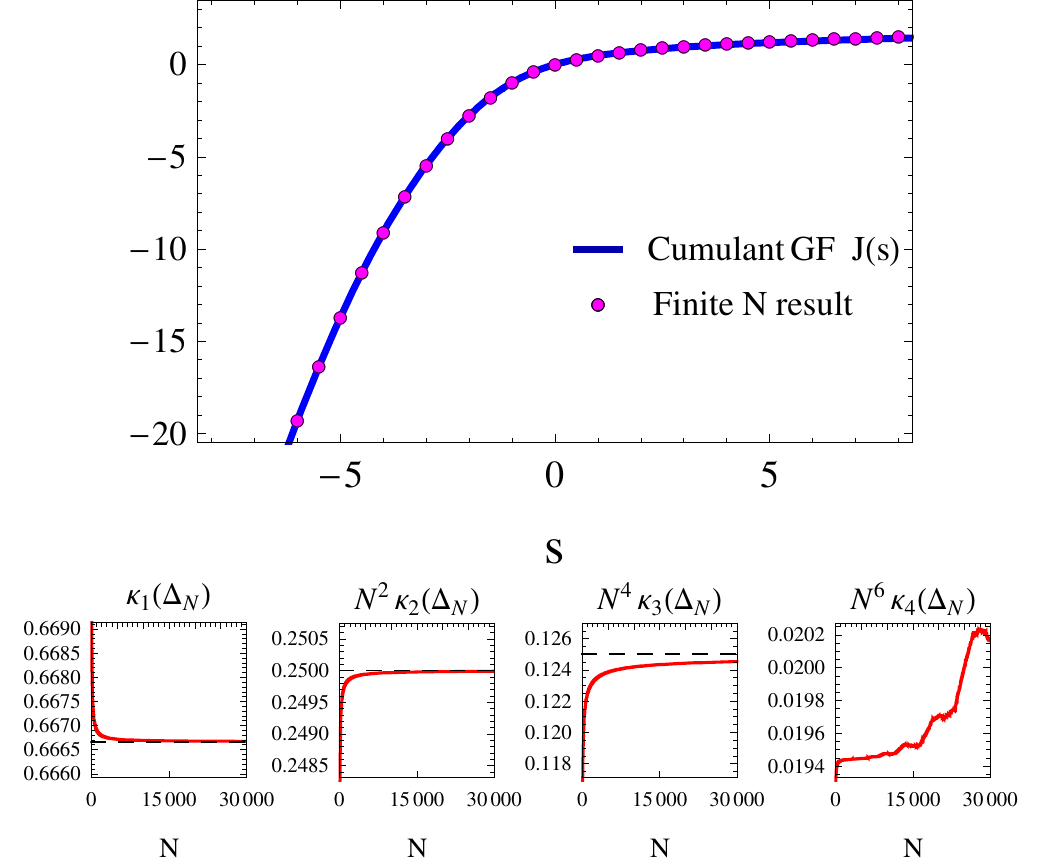}
\caption{(color online) Top: The cumulant generating function $J(s)$ in~\eqref{eq:result_J} (solid line) compared with the exact finite $N$ formula~\eqref{eq:finiteN} (at $\beta=2$) for $N=16$ (dots).  Already for such a small value of $N$, the large deviation function $J(s)$ matches the finite $N$ expression of $-1/(2N^{2})\log\widehat{\mathcal{P}}_{N,\beta=2}(s)$ with an error $\Ord(10^{-2})$ in the plotted region. Bottom: Rescaled cumulants $N^{2(r-1)}\kappa_m(\Delta_N)$ for $m=1,2,3,4$ at increasing values of $N$ for $\beta=2$. The first three cumulants have a regular behaviour in $N$ and approach their leading order~\eqref{eq:cumulants}, while $N^6\kappa_4$ behaves irregularly and does not converge for large $N$.}
\label{fig:J}
\end{figure}

\textit{The integrable case $\beta=2$.}
As already mentioned, the value $\beta=2$ is special. At this value of the plasma parameter the 2D-OCP canonical measure has a determinantal structure and the partition function is integrable at finite $N$. This fact has been largely exploited in random matrix theory. For $\beta=2$, the Laplace transform of radially symmetric linear statistics has been computed exactly by Forrester~\cite{Forrester99}. By specializing his result to $\Delta_N$ we obtain the following finite $N$ formula at (crucially) $\beta=2$: 
\be
\widehat{\mathcal{P}}_{N,\beta=2}(s)=\prod_{\ell=1}^N\left[\frac{\int_0^{\infty}\de t\,e^{-(t+2s\sqrt{Nt})}t^{\ell-1}}{\int_0^{\infty}\de t\,e^{-t}t^{\ell-1}}\right]. \label{eq:finiteN}
\ee
In Fig.~\ref{fig:J} we show that our large $N$ result~\eqref{eq:result_J} agrees with this finite $N$ result even for moderate values of $N$. We stress however that the large deviation formula is valid for any value of $\beta>0$. Moreover, the route presented here shows explicitly that, even at $\beta=2$, the function $-1/(\beta N^2)\log\widehat{\mathcal{P}}_{N,\beta}(s)$ analytic at finite $N$ (see~\eqref{eq:finiteN}) develops a non-analyticity as a result of the thermodynamic limit. To understand better the emergence of the singularity we have performed another numerical experiment. For $\beta=2$ the (unordered!) collection of displacements $\{r_k\}$ of the charges is distributed as a collection of \emph{independent} random variables $\{N^{-1/2}\xi_k\}$ where $\xi_k\geq0$ has density $f_k(x)=(2/\Gamma(k))x^{2k-1}
\exp(-x^2)$ and  moments $\avg{\xi_k^{m}}=\Gamma(k+m/2)/\Gamma(k)$~\cite{Kostlan92,Rider03}. The cumulants $\kappa_m$ ($m\geq1$) are linear functionals for independent variables and hence we have $\kappa_m(\Delta_N)=N^{-\frac{3m}{2}}\sum_{k=1}^N\kappa_m(\xi_k)$. We have evaluated numerically this sum for increasing values of $N$ and compared the output with our $N\gg1$ results on the cumulants~\eqref{eq:cumulants} (see Fig.~\ref{fig:J}).  For $m=1,2,3$, the sum asymptotically approaches~\eqref{eq:cumulants}, while the fourth cumulant $\kappa_4(\Delta_N)$ shows a bizarre behaviour for large $N$. For $\beta=2$, this is a further confirmation of the weak singularity of the large deviations functions.

\textit{Role of the  logarithmic interaction.}
In order to convey  the key ideas of this paper, we conclude with a brief digression to clarify the role of the 2D Coulomb interaction in the mechanism of the phase transition.
The computation of $J(s)$ led to study the minimisation problem~\eqref{eq:min}. For $s\geq0$ the effective external potential $V_s(r)$ in~\eqref{eq:potential_s} is convex and the distribution of charges $\mu^{\star}_s$ is supported on a disk. When $s$ is negative the convexity of $V_s(r)$ is broken and the plasma distribution concentrates on an annulus. One might think that this change in the potential  is  sufficient to modify the topology of the charge distribution and to induce a phase transition even in absence of interaction between the charges. This is incorrect, as we will demonstrate.  

Removing the logarithmic interaction from~\eqref{eq:hamiltonian} turns the plasma into a non-interacting gas with single-particle energy $H\left(\left\{\vec{r}_k\right\}\right)=(1/2)\sum_kr_k^2$ at inverse temperature $\beta>0$~\footnote{Here we drop off the $N$ scaling in front of the harmonic potential in order to keep the particles of the gas $r_{k}=\Ord(1)$.}.
Since the particles are independent, the single-particle distribution is Gaussian (the system is no more confined in a bounded region) and $\Delta_N=N^{-1}\sum_k r_k$ is a sum of independent and identically distributed displacements. 
Hence, an elementary computation shows that $\langle\Delta_N\rangle=\sqrt{\pi/(2\beta)}$,  $\mathrm{var}(\Delta_N)=(4-\pi)/(2\beta N)$ and that the Laplace transform is $\widehat{\mathcal{P}}_{N,\beta}(s)=\langle e^{-\beta sN\Delta_N}\rangle=[1-\sqrt{2\pi\beta}\,s\exp(-\beta s^2/2)\Phi(-\beta s)]^N$, where $\Phi(x)$ is the standard Gaussian cumulative function. (Note the different scaling in $N$ and $\beta$-dependence in the above expressions compared with the 2D-OCP.) Therefore, in absence of interaction the free energy of the system is analytic in $s$: non-interacting particles do not undergo phase transitions.

\textit{Conclusions.} Many of the ideas presented here can be generalized to other problems. In particular, the method to compute large deviations formulas for 2D-OCP is applicable to a large class of radially symmetric observables $A_N=\sum_i f(|\vec{r_i}|)$ with $f(r)$ continuous. The problem of discontinuous statistics, like the charge fluctuations (i.e. the index problem in random matrices), is delicate and needs more refined tools (see~\cite{Forrester14} for recent developments). 

We detect a novel and unusual type of weak phase transition due to a change of topology in the plasma distribution.
Therefore this work naturally prompts the following question: is this fourth-order phase transition on the plane as ubiquitous and universal as the third-order transition for charged particles on the line? 
We mention that similar disk-annulus phase transitions have been detected for non-Hermitian non-Gaussian random matrices~\cite{Feinberg97,Feinberg01}. However, non-Gaussian ensembles do not have a 2D-OCP interpretation and it is not clear how to relate these works with our findings.

An interesting question is to look at the finite $N$ corrections of the large deviations principle~\eqref{eq:J}-\eqref{eq:rate_funct_def}. At microscopic scales it has been shown~\cite{Sandier,Choquard83,Sari76a} that, among all lattice configurations, the charges of the plasma prefer to arrange themselves in a triangular one (the very same of the \emph{Abrikosov lattice}~\cite{Abrikosov57} of vortices in the Ginzburg-Landau model). Showing that the triangular lattice is the global minimizer of the 2D-OCP energy requires a careful analysis beyond the mean-field limit. Such a conjecture is still unsolved even in the integrable case $\beta=2$.

\textit{Acknowledgments.}  We thank Pierpaolo Vivo for helpful suggestions. This work was partially supported by EPSRC Grant number EP/L010305/1. FDC acknowledges the partial support of Gruppo Nazionale di Fisica Matematica GNFM-INdAM. AM acknowledges the support of the Leverhulme Trust Early Career
Fellowship (ECF 2013-613). No empirical or experimental data were created during this study.



\begin{thebibliography}{99}
\bibitem{Sandier} E. Sandier and S. Serfaty,
\emph{From the Ginzburg-Landau Model to Vortex Lattice Problems},
Commun. Math. Phys. {\bf 313}, 635–743 (2012).

\bibitem{GL} V.~L. Ginzburg, L.~D. Landau, 
\emph{Collected papers of L.~D.Landau}. Edited by D. Ter.
Haar, Pergamon Press, Oxford (1965).

\bibitem{Correggi08} M. Correggi and J. Yngvason,
\emph{Energy and vorticity in fast rotating Bose-Einstein condensates},
J. Phys. A: Math. Theor. {\bf 41}, 445002 (2008).

\bibitem{Correggi12} M. Correggi, F. Pinsker, N. Rougerie and J. Yngvason,
\emph{Critical rotational speeds for superfluids in homogeneous traps},
J. Math. Phys. {\bf 53}, 095203 (2012).

\bibitem{Dyson62}
F.~J. Dyson,
\emph{A Brownian Motion Model for the Eigenvalues of a Random Matrix},
J. Math. Phys. {\bf 3}, 1191 (1962).

\bibitem{Osada13} H. Osada,
\emph{Interacting Brownian motions in infinite dimensions with logarithmic interaction potentials},
Ann. Prob. {\bf 41}, 1-49 (2013).

\bibitem{Ginibre} J. Ginibre,
\emph{Statistical ensembles of complex, quaternion, and real matrices}, 
J. Math. Phys. {\bf 6}, 440-449 (1965).

\bibitem{Deutsch79} C. Deutsch, H. E. Dewitt and Y. Furutani,
\emph{Debye thermodynamics for the two-dimensional one-component plasma},
Phys. Rev. A {\bf 20}, 2631 (1979).

\bibitem{Jancovici81} B. Jancovici,
\emph{Exact Results for the Two-Dimensional One-Component Plasma},
Phys. Rev. Lett. {\bf 46}, 386 (1981).

\bibitem{Alastuey81} A. Alastuey and B. Jancovici,
\emph{On the classical two-dimensional one-component Coulomb plasma},
J. Physique {\bf 42}, 1-12 (1981).

\bibitem{Johannesen83} S. Johannesen and D. Merlini,
\emph{On the thermodynamics of the two-dimensional jellium}
J. Phys. A: Math. Gen. {\bf 16}, 1449-1463 (1983).

\bibitem{Forrester85} P.~J. Forrester,
\emph{The two-dimensional one-component plasma at $\Gamma=2$: metallic boundary},
 J. Phys. A: Math. Gen. {\bf 18}, 1419 (1985).

\bibitem{Cornu88} F. Cornu and B. Jancovici,
\emph{Two- Dimensional Coulomb Systems: a Larger Class of Solvable Models},
Europhys. Lett. {\bf 5}, 125-128 (1988).


\bibitem{Forrester98} P.~J. Forrester,
\emph{Exact results for two-dimensional Coulomb systems},
Phys. Rep. {\bf 301}, 235270 (1998).

\bibitem{Samaj03} L. \v{S}amaj,
\emph{The statistical mechanics of the classical two-dimensional Coulomb gas is exactly solved},
J. Phys. A: Math. Gen. {\bf 36}, 5913 (2003). 

\bibitem{Fischmann11}
J. Fischmann and P.~J. Forrester,
\emph{One-component plasma on a spherical annulus and a random matrix ensemble},
J. Stat. Mech. {\bf P10003},  (2011).

\bibitem{Choquard83} Ph. Choquard and J. Clerouin,
\emph{Cooperative Phenomena below Melting of the One-Component Two-Dimensional Plasma},
Phys. Rev. Lett. {\bf 50}, 2086 (1983).

\bibitem{Dean06} D.~S. Dean and S.~N. Majumdar, 
\emph{Large Deviations of Extreme Eigenvalues of Random Matrices},
Phys. Rev. Lett. {\bf 97}, 160201 (2006);
\emph{Extreme value statistics of eigenvalues of Gaussian random matrices}, 
Phys. Rev. E {\bf 77}, 041108 (2008).

\bibitem{Facchi08} P. Facchi, U. Marzolino, G. Parisi, S. Pascazio and A. Scardicchio, 
\emph{Phase Transitions of Bipartite Entanglement},
Phys. Rev. Lett. {\bf 101}, 050502 (2008).

\bibitem{Vivo08} P. Vivo, S.~N. Majumdar and O. Bohigas, 
\emph{Distributions of Conductance and Shot Noise and Associated Phase Transitions}, 
Phys. Rev. Lett. {\bf101}(21), 216809 (2008);
\emph{Probability distributions of linear statistics in chaotic cavities and associated phase transitions}, 
Phys. Rev. B {\bf 81}, 104202 (2010).

\bibitem{DePasquale10} A. De Pasquale, P. Facchi, G. Parisi, S. Pascazio and A. Scardicchio, 
\emph{Phase transitions and metastability in the distribution of the bipartite entanglement of a large quantum system},
Phys. Rev. A {\bf 81}, 052324 (2009).

\bibitem{Majumdar11} S.~N. Majumdar, C. Nadal, A. Scardicchio and P. Vivo,
\emph{Index distribution of Gaussian Random Matrices},
Phys. Rev. Lett. {\bf 103}, 220603 (2009);
\emph{How many eigenvalues of a Gaussian random matrix are positive?},
Phys. Rev. E {\bf 83}, 041105 (2011).

\bibitem{Borot12} G. Borot and C. Nadal,
\emph{Purity distribution for generalized random Bures mixed states},
J. Phys. A: Math. Theor. {\bf 45}, 075209 (2012).

\bibitem{Facchi13} P. Facchi, G. Florio, G. Parisi, S. Pascazio and K. Yuasa, 
\emph{Entropy-driven phase transitions of entanglement},
Phys. Rev. A {\bf 87}, 052324 (2013).

\bibitem{Texier13}
C. Texier and  S.~N. Majumdar,
\emph{Wigner time-delay distribution in chaotic cavities and freezing transition},
Phys. Rev. Lett. {\bf 110}, 250602 (2013)~;
\textit{ibid} {\bf 112}, 139902(E) (2014).

\bibitem{Majumdar14}
S.~N. Majumdar and G. Schehr,
\emph{Top eigenvalue of a random matrix: large deviations and third order phase transition}, 
J. Stat. Mech. {\bf P01012} (2014).

\bibitem{Jancovici93} B. Jancovici. J.~L. Lebowitz, G. Manificat, 
\emph{Large Charge Fluctuations in Classical Coulomb Systems},
J. Stat. Phys. {\bf 72} 3/4 (1993).

\bibitem{Allez14} R. Allez, J. Touboul and G. Wainrib,
\emph{Index distribution of the Ginibre ensemble}, 
J. Phys. A: Math Theor. {\bf 47}, 042001 (2014).

\bibitem{Arous98} G.~B. Arous and O. Zeitouni,
\emph{Large deviations from the circular law},
ESAIM Probab. Statist. {\bf 2}, 123-134 (1998).

\bibitem{Petz98} D. Petz and F. Hiai,
\emph{Logarithmic energy as an entropy functional},
Contemp. Math. AMS {\bf 217}, 205-221 (1998).


\bibitem{Gartner} J. G\"artner,
 \emph{On large deviations from the invariant measure}, 
Theory Probab. Appl. {\bf 22}, 24-39 (1977).

\bibitem{Ellisthm} R. S. Ellis, 
\emph{Large deviations for a general class of random vectors}, 
Ann. Probab. {\bf 12} (1), 1-2 (1984).

\bibitem{Forrester99} P.~J. Forrester, 
\emph{Fluctuation formula for complex random matrices},
J. Phys. A: Math. Gen. {\bf 32}, L159-L163 (1999).

\bibitem{Saff} E.~B. Saff and V. Totik,
\emph{Logarithmic Potentials with External Fields}, 
Springer-Verlag Berlin Heidelberg GmbH (1991).

\bibitem{Girko84} V.~L. Girko,
\emph{Circular law},
Teor. Veroyatnost. i Primenen. {\bf29}, 669 (1984).
 	
\bibitem{Bai97} Z.~D. Bai,
\emph{Circular law}, 
Ann. Prob. {\bf 25}, 494-529 (1997).

\bibitem{Bordenave12} C. Bordenave and D. Chafa\"{i},
\emph{Around the circular law}, 
Probability Surveys {\bf 9}, 1-89 (2012).

\bibitem{Sari76} R.~R. Sari  and D. Merlini,
\emph{On the $\nu$-Dimensional One-Component Classical Plasma: The Thermodynamic Limit Problem Revisited},
J. Stat. Phys. {\bf 14}, 91 (1976).

\bibitem{Sari76a} R.~R. Sari, D. Merlini and R. Carlinon,
\emph{On the ground state of the one-component classical plasma},
J. Phys. A: Math. Gen. {\bf 9}, 1539 (1976).


\bibitem{Kostlan92} E. Kostlan,
\emph{On the Spectra of Gaussian Matrices},
Linear Algebra Appl. {\bf 162-164}, 385-388 (1992).

\bibitem{Rider03} B. Rider,
\emph{A limit theorem at the edge of a non-Hermitian random matrix ensemble},
J. Phys. A {\bf 36}, 3401-3409 (2003).

\bibitem{Forrester14}  T. Can, P. J. Forrester, G. T\'ellez, P. Wiegmann,
\emph{Exact and Asymptotic Features of the Edge Density
Profile for the One Component Plasma in Two
Dimensions},
J. Stat. Phys. {\bf 158}, 1147-1180 (2015).


\bibitem{Feinberg97} J. Feinberg and A. Zee,
\emph{Non-gaussian non-hermitian random matrix theory: Phase transitions and addition formalism},
Nucl. Phys. B {\bf 501}, 643-669 (1997).

\bibitem{Feinberg01} J. Feinberg, R. Scalettar and A. Zee,
\emph{``Single ring theorem'' and the disk-annulus phase transition},
J. Math. Phys. {\bf 42}, 5718-5740 (2001).

\bibitem{Abrikosov57} A. Abrikosov, 
\emph{On the magnetic properties of superconductors of the second group},
Soviet Phys. JETP {\bf 5}, 1174–1182 (1957).


\bibitem{Ellis99} R.~S. Ellis, 
\emph{The theory of large deviations: from Boltzmann's 1877 calculation to equilibrium macrostates in 2D turbulence},
Physica D {\bf 133}, 106-136 (1999).

\bibitem{Touchette09} 
H. Touchette, 
\emph{The large deviation approach to statistical mechanics}, 
Phys. Rep. {\bf 478}, 1 (2009).



\end{thebibliography}
\end{document}